\documentclass[aps,prl,twocolumn,groupedaddress]{revtex4-2}

\usepackage{color}

\usepackage[utf8]{inputenc}
\usepackage[english]{babel}
\usepackage{amsmath,graphicx,enumerate}

\begin{document}

\title{Entropons as collective excitations in active solids} 


\author{L. Caprini$^{1}$}
\author{U. Marini Bettolo Marconi$^{2,3}$}
\author{A. Puglisi$^{4,5}$} 
\author{H. L\"owen$^{1}$}

\affiliation{$^1$ Heinrich-Heine-Universit\"at D\"usseldorf, D\"usseldorf, Germany.\\ 
$^2$ Scuola di Scienze e Tecnologie, Universit\`a di Camerino - via Madonna delle Carceri, 62032, Camerino, Italy.\\
$^3$ Istituto Nazionale di Fisica Nucleare, Sezione di Perugia, Via A. Pascoli, I-06123 Perugia, Italy. 
\\ $^4$ Istituto dei Sistemi Complessi - CNR and 
Universit\`a di Roma Sapienza, P.le Aldo Moro 2, 00185, Rome, Italy 
\\ $^5$ INFN, University of Rome Tor Vergata, Via della Ricerca Scientifica 1, 00133, Rome, Italy}

\date{\today}


\begin{abstract}

The vibrational dynamics of solids is described by phonons constituting basic collective excitations in equilibrium crystals.
Here we consider an active crystal composed of self-propelled particles which bring the system into a non-equilibrium steady-state governed by entropy production.
Calculating the entropy production spectrum, we put forward the picture of ``entropons'', which 
are vibrational collective excitations responsible for entropy production. Entropons are purely generated by activity
and coexist with phonons but dominate over them for large self-propulsion strength.
The existence of entropons can be verified in experiments on dense self-propelled colloidal Janus-particles and granular active matter, as well as in living systems such as dense cell monolayers.
\end{abstract}

\maketitle


Active matter~\cite{marchetti2013hydrodynamics, elgeti2015physics} includes a broad range of systems composed of particles that locally convert energy from the environment into directed motion~\cite{bechinger2016active, gompper20202020}.
The energy exchange with the environment, often induced by chemical reactions or 
self-imposed gradients, leads to self-propulsion of the particles and, thus, drives an active
system intrinsically out of equilibrium. 

Dense systems of self-propelled particles  are rather ubiquitous in nature and often form crystalline structures.
Cells monolayers of the human body~\cite{alert2020physical, garcia2015physics, yang2017correlating, henkes2020dense} and dense colonies of bacteria~\cite{peruani2012collective, petroff2015fast, wioland2016ferromagnetic} which populate the human skin are common examples.
Moreover, active colloidal Janus particles may cluster and form dense crystallites~\cite{buttinoni2013dynamical}, named ``living crystals''~\cite{palacci2013living, mognetti2013living}.
These active crystals show fascinating phenomena uncommon for equilibrium solids~\cite{menzel2013unidirectional, ferrante2013elasticity, lin2021order, ophaus2021two} ranging from  ``traveling''  crystals~\cite{menzel2013traveling, praetorius2018active, briand2018spontaneously}, intrinsic velocity correlations~\cite{szamel2015glassy, caprini2021spatial, szamel2021long, kuroda2022anomalous} and spontaneous velocity alignment~\cite{caprini2020spontaneous, caprini2020hidden} to collective rotations~\cite{ferrante2013elasticity, huang2020dynamical}.
Activity also shifts the equilibrium  freezing transition significantly~\cite{bialke2012crystallization, digregorio2018full, klamser2018thermodynamic, omar2021phase} and affects the nature of the two-dimensional melting transition~\cite{digregorio2018full, pasupalak2020hexatic, li2021melting, negro2022inertial, hopkins2022yield}.

External forces or internal mechanisms that dissipate energy drive a system away from  equilibrium and spontaneously produce entropy. While self-propulsion is generated by an uptake of energy from the environment, likewise active particles dissipate energy back into the environment. 
The resulting energy conversion leads to local entropy production~\cite{speck2016stochastic, pigolotti2017generic, mandal2017entropy, caprini2018comment, szamel2019stochastic}. 
Quantifying the non-equilibrium character of active systems via entropy production has represented a topic of central interest in recent years~\cite{dabelow2019irreversibility, caprini2019entropy, fodor2021irreversibility, o2022time}.
This topic has been investigated numerically, simulating both active field theories~\cite{nardini2017entropy, borthne2020time, paoluzzi2022scaling, cates2022stochastic}, active particle dynamics in interacting systems~\cite{crosato2019irreversibility, chiarantoni2020work, grandpre2021entropy}, as well as colloids in the presence of an active bath~\cite{pietzonka2017entropy, chaki2018entropy}.
Particular attention has been devoted to phase-separated configurations where the main contribution to the spatial profile of the entropy production has been observed at the interface between dense and dilute phases~\cite{nardini2017entropy, martin2021statistical, guo2021play}. Conversely, analytical results for entropy production have been only derived for simple cases, such as the potential-free particle~\cite{shankar2018hidden, razin2020entropy, cocconi2020entropy}, and for near-equilibrium regimes through perturbative methods~\cite{fodor2016far, marconi2017heat}. 
Entropy production in active crystals, however, remains  unexplored. 

\begin{figure}[!t]
\centering
\includegraphics[width=1\linewidth,keepaspectratio]{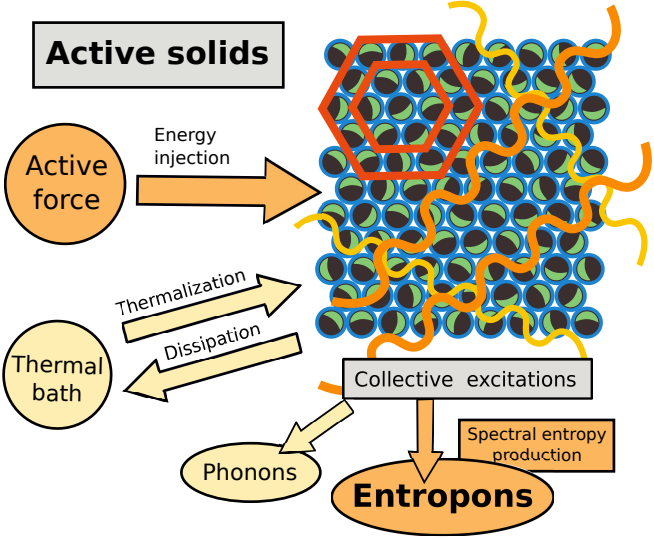}
\caption{
Schematic representation of a solid formed by self-propelled particles that locally inject energy via their active force. Each  particle is represented by a capped sphere. The orientation of the green hemisphere denotes the direction of the active force, while red hexagons are drawn to highlight the hexagonal crystal structure.
Undulated curves on the solid are schematic illustrations of the vibrational excitations: the phonons (yellow) in both
 equilibrium and non-equilibrium, and the entropons (orange) in non-equilibrium. 
}\label{fig:Fig0}
\end{figure}

In this Letter, we fill this gap and calculate the entropy production in an active crystal.
Performing a systematic spectral analysis, we show the existence of novel collective vibrational excitations in non-equilibrium active solids which are responsible for entropy production.
Following the standard nomenclature of solid-state physics~\cite{pines2018elementary}, we term these new modes ``entropons''. Unlike phonons, which describe the vibrational dynamics of equilibrium solids, entropons exist only in non-equilibrium, i.e. they are purely induced by activity. Entropons coexist with phonons but dominate over them for large activity and therefore represent the thermodynamically most relevant modes of an active crystal far from equilibrium.
The underlying basic picture is shown in Fig.~\ref{fig:Fig0}: in equilibrium, the thermal bath coupled to a crystal induces collective vibrational excitations like phonons (yellow color in Fig.~\ref{fig:Fig0}); in non-equilibrium, the active force, acting on each particle and producing their self-propulsion, injects energy into the crystal, this energy is dissipated in the environment, and the system produces entropy.
In this process, entropons are generated as new collective vibrations, encoding the entropy production (orange color in Fig.~\ref{fig:Fig0}).
Our analysis is based on analytical theory combined with particle-resolved computer simulations and can in principle be verified in experiments with highly packed self-propelled colloidal Janus-particles~\cite{klongvessa2019nonmonotonic} or vibrated granular grains~\cite{dauchot2005dynamical, puglisi2012structure, briand2016crystallization}, as well as in living systems such as confluent cell monolayers~\cite{garcia2015physics, henkes2020dense}.


We study a two-dimensional crystal of $N$ inertial active Brownian particles, in a square box of size $L$ with periodic boundary conditions. 
Each particle with mass $m$ evolves through an underdamped dynamics~\cite{takatori2017inertial, mandal2019motility, leoni2020surfing, caprini2021inertial} for its position, $\mathbf{x}_i$ and velocity $\mathbf{v}_i=\dot{\mathbf{x}}_i$
\begin{subequations}
\label{eq:activedynamics}
\begin{align}
&m\dot{\mathbf{v}}_i= -\gamma \mathbf{v}_i + \mathbf{F}_i +\sqrt{2 T \gamma} \boldsymbol{\zeta}_i + \mathbf{f}^a_i\\
&\dot{\theta}_i=\sqrt{2D_r} \eta_i \,,
\end{align}
\end{subequations}
where  $\boldsymbol{\zeta}_i$ and $\eta_i$ are Gaussian white noises with zero average and unit variance.
The term $\mathbf{f}^a_i=\gamma v_0 \mathbf{n}_i$ models the active force, $v_0$ being the swim velocity and $\mathbf{n}_i=(\cos\theta_i, \sin\theta_i)$ the orientational unit vector, determined by an orientational angle $\theta_i$.
The coefficients $\gamma$ and $T$ correspond to the friction coefficient and the temperature of the solvent bath, respectively, and define the inertial time $\tau_I=m/\gamma$. 
$D_r$ is the rotational diffusion coefficient which determines the persistence time, $\tau=1/D_r$, i.e. the time that the particle needs to randomize its orientation. The single-particle dynamics is often described in terms of the so-called active temperature, $T_a=v_0^2 \tau \gamma$ that vanishes in the equilibrium limits, either $\tau \to 0$ or $v_0\to0$.
The interaction force $\mathbf{F}_i$ stems from  a soft repulsive pair potential, $U_{\text{tot}}=\sum_i U(|\mathbf{x}_{i}-\mathbf{x}_{i}|)$, where $U=4\epsilon [(d_0/r)^{12}-(d_0/r)^{6}]$ if $r<2^{1/6}$ and zero otherwise (WCA potential), $\epsilon$ and $d_0$ being the energy scale and the particle diameter, respectively. 
The packing fraction $\phi=\rho d_0^2\pi/4$ (where $\rho=N/L^2$ denotes the number density), is chosen high enough to ensure a solid-like behavior characterized by a defect-free triangular lattice as illustrated in Fig.~\ref{fig:Fig0}. Further details are reported in the Supplemental Material (SM).
%


The non-equilibrium properties of the system are investigated by applying path-integral techniques to calculate the total entropy production, $\dot{s}=\lim_{t\to\infty}\langle\log \left(\mathcal{P}/\mathcal{P}_r \right) \rangle/t$, where $\mathcal{P}$ and $\mathcal{P}_r$ represent the path probabilities of the forward and backward trajectory~\cite{dabelow2019irreversibility, caprini2019entropy}, respectively, see SM for details and Ref.~\cite{seifert2012stochastic} for a general review.
The steady-state entropy production can be decomposed into its space-time Fourier spectrum as
\begin{equation}
\label{eq:def_spectralentropyprod}
\dot{s}=\int_{\Omega} \frac{d {\mathbf{q}}}{\Omega} \int_{-\infty}^{\infty} \frac{d\omega}{2\pi} \, \sigma(\omega, \mathbf{q})
\end{equation}
where $\mathbf{q}$ is the wave vector, $\omega$ the frequency and $\Omega$ represents the area of the first two-dimensional Brillouin zone.

As a main result of this Letter, $\sigma(\omega, \mathbf{q})$ is analytically predicted as (see SM)
\begin{flalign}
\label{eq:spectral_entropyprod}
\sigma(\omega, \mathbf{q})&=\lim_{t \to \infty} \frac{1}{t}\frac{1}{2T}\left(\langle \tilde{\mathbf{v}}( \omega,\mathbf{q}) \cdot \tilde{\mathbf{f}}_a(-\omega,-\mathbf{q}) \rangle + \text{CC}\right) \nonumber\\
&\approx \frac{T_a}{T} \frac{K(\omega)}{\tau_I^2} \frac{\tau_I^2\omega^2}{\tau_I^2(\omega^2-\omega^2(\mathbf{q}))^2+\omega^2}\,,
\end{flalign}
where the symbol $\text{CC}$ stands for complex conjugate.
The vectors $\tilde{\mathbf{f}}_a(- \omega,-\mathbf{q})$ and $\tilde{\mathbf{v}}(\omega,\mathbf{q})$ are the Fourier transforms of active force and velocity fields in the frequency and wave vector domains (see SM for their definitions).
The second line of Eq.~\eqref{eq:spectral_entropyprod} is obtained in the limit of a harmonic crystal and expresses $\sigma$ as a function of the parameters of the model since the shape function $K(\omega)$ explicitly reads
\begin{equation}
K(\omega)=\frac{1}{1+\omega^2\tau^2} \,.
\end{equation}
The term $\omega^2(\mathbf{q})$ in Eq.~\eqref{eq:spectral_entropyprod} denotes the phonon dispersion relation of equilibrium solids, whose expression is reported in the SM for a triangular lattice. 
In general, $\omega^2(\mathbf{q}) \propto\omega_E$ where $\omega_E=\frac{1}{2m}\left( U''(\bar{x})+\frac{U'(\bar{x})}{\bar{x}}\right)$ is the Einstein frequency of the solid determined by the derivative of the potential calculated at the average distance between neighboring particles, $\bar{x} \sim 1/\sqrt{\rho}$.

The dynamical correlation function, $\left\langle {\tilde{\mathbf{u}}}(\omega,\mathbf{q})\cdot {\tilde{\mathbf{u}}}(-\omega,-\mathbf{q})  \right\rangle$, where $\tilde{\mathbf{u}}$ is the Fourier transform of the particle displacement with respect to the unperturbed position of its lattice, can be expressed in terms of $\sigma$ as (see SM)
\begin{equation}
\label{eq:harada_sasa}
\frac{1}{T} \left\langle {\tilde{\mathbf{u}}}(\omega,\mathbf{q})\cdot {\tilde{\mathbf{u}}}( -\omega,-\mathbf{q})  \right\rangle  = \frac{\text{Im}[ \mathcal{R}_{uu}( \omega,\mathbf{q})]}{\omega}
+ \frac{\sigma(\omega,\mathbf{q})}{\omega^2\gamma} 
\end{equation}
where $\text{Im}[ \mathcal{R}_{uu}(\omega,\mathbf{q})]$ is the imaginary part of the response function due to a small perturbation, $h$, evaluated in the frequency and wave vector domains.
The response is defined as $\mathcal{R}_{uu}(\omega,\mathbf{q})=\delta \langle \hat{\mathbf{u}}(\omega)\rangle/\delta h(\omega)|_{h=0}$ and one  has
\begin{equation}
\text{Im}[ \mathcal{R}_{uu}(\mathbf{q}, \omega)]=\frac{\omega \tau_I}{\tau_I^2\left(\omega^2-\omega^2(\mathbf{q})\right)^2+\omega^2} \,.
\end{equation}
The relation~\eqref{eq:harada_sasa} indicates that $\left\langle {\tilde{\mathbf{u}}}(\omega,\mathbf{q})\cdot {\tilde{\mathbf{u}}}( -\omega,-\mathbf{q})  \right\rangle$ is the sum of two terms corresponding to: i) thermally excited crystal vibrations, independent of activity
that we identify with phonons and ii) additional vibrational contributions of the solid associated with the entropy production that are purely induced by the active force.
We remark that the latter are dominant if $T_a =v_0^2\tau\gamma \gg T$, i.e. far from equilibrium: There exists a typical $\tau$ (or $v_0$) at which their contribution becomes negligible with respect to the one of the phonons. 
In the limit of zero active force ($T_a \to 0$), the response balances the l.h.s of Eq.~\eqref{eq:harada_sasa} and the entropy production vanishes as equilibrium imposes. 

\begin{figure}[!t]
\centering
\includegraphics[width=1\linewidth,keepaspectratio]{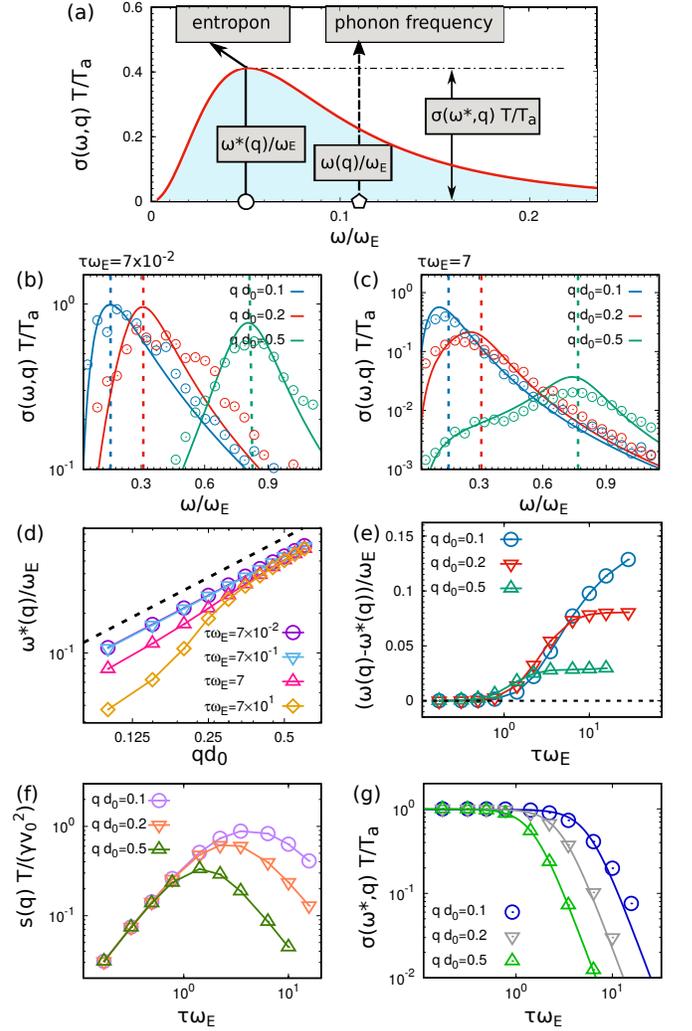}
\caption{Spectral entropy production. 
(a): schematic representation of the spectral entropy production $\sigma(\omega, \mathbf{q}) T/T_a$ to identify entropons as a peak in the spectrum.
(b) and~(c):  $\sigma(\omega, \mathbf{q}) T/T_a$ as a function of $\omega/\omega_E$ for different values of the rescaled wave vector $q d_0$ for $\tau\omega_E=7\times10^{-2}, 7$, respectively.
(d): frequency, $\omega^*/\omega_E$ where $\sigma(\omega, \mathbf{q})$ is peaked as a function of $q d_0$ for different value of the rescaled persistence time $\tau\omega_E$. The black dotted line is an eye guide to show a linear curve.
(e): difference between the frequency of the phonon, $\omega(\mathbf{q}^*)T/T_a$, and $\omega^*(\mathbf{q}^*)T/T_a$ as a function of $\tau \omega_E=\omega_E/D_r$ for different values of $q d_0$.
(f): Integrated entropy production, $s(\mathbf{q}) T/(v_0^2 \gamma)$ as a function of $\tau \omega_E$ for different values of $q d_0$.
(g): maximal value of the entropy production, $\sigma(\omega^*, \mathbf{q}) T/T_a$, as a function of $\tau \omega_E$ for different values of $q d_0$. Lines are obtained by fitting the function $1/(1+b\tau^2\omega_E^2)$, where $b$ is a fitting parameter.
}\label{fig:Fig1}
\end{figure}

A typical shape of $\sigma(\omega, \mathbf{q})T/T_a$ is shown in Fig.~\ref{fig:Fig1}~(a) as a function of $\omega/\omega_E$ for a given $\mathbf{q}$.
A sharp peak occurs at a characteristic frequency $\omega^*(\mathbf{q})$.
We identify this peak with an elementary excitation in the crystal and coin the term ``entropon'' to describe it following the standard notation of elementary excitations in solids~\cite{pines2018elementary}: Each entropon is identified with a peak of ${\sigma}(\omega, \mathbf{q})$.

Fig.~\ref{fig:Fig1}~(b) and~(c) show $\sigma(\omega, \mathbf{q})T/T_a$ as a function of $\omega/\omega_E$ for different values of $\mathbf{q}$ revealing a good agreement between numerical simulations and theory, Eq.\eqref{eq:spectral_entropyprod}.
Close to the equilibrium, in the regime of small persistence time such that $\tau=1/D_r \ll 1/\omega_E$ (Fig.~\ref{fig:Fig1}~(b)),
the peaks of $\sigma(\omega, \mathbf{q})T/T_a$ occur at the phonon frequency $\omega^*(\mathbf{q})=\omega(\mathbf{q})$.
In this regime, entropons have the same properties of phonons but their amplitudes are small and proportional to $\tau$ (because of the prefactor $T_a$). In these conditions, the active force behaves as an additional thermal source at effective temperature $T_a$.
In the opposite regime of large persistence time, $\tau=1/D_r \gg 1/\omega_E$, (Fig.~\ref{fig:Fig1}~(c)), the peaks of $\sigma(\omega, \mathbf{q})$ are shifted to $\omega^*(\mathbf{q}) < \omega(\mathbf{q})$. 
As a consequence, entropons are different from phonons since the crystal vibrations are now peaked at frequencies 
not coinciding with those typical of equilibrium solids.
The frequency $\omega^*(\mathbf{q})$ which maximizes $\sigma(\omega, \mathbf{q})$ is reported in Fig.~\ref{fig:Fig1}~(d) as a function of $\mathbf{q}$ for different values of the rescaled persistence time, $\tau\omega_E$, while the difference $\omega(\mathbf{q})- \omega^*(\mathbf{q})$ is shown in Fig.~\ref{fig:Fig1}~(e) as a function of $\tau \omega_E$ for different values of $\mathbf{q}$.
Despite $\omega^*(\mathbf{q})$ linearly increases with $\mathbf{q}$ in the small persistence regime, a clear discrepancy from the linear law emerges in the large persistence regime for small $\mathbf{q}$, where entropons follow a non-linear dispersion relation. 
As a consequence, the difference $\omega(\mathbf{q})- \omega^*(\mathbf{q})$ grows when $\tau \omega_E$ is increased much more as $\mathbf{q}$ is decreased.

Fig.~\ref{fig:Fig1}~(f) displays the integral over $\omega$ of the spectral entropy production, $s(\mathbf{q})=\int d\omega \sigma(\omega, \mathbf{q})=\rho\frac{T_a}{T }(\tau_I+\tau)^{-1}(1+ \frac{\tau^2\tau_I}{\tau+\tau_I} \omega^2(\mathbf{q}))^{-1}$, as a function of $\tau\omega_E$ for different $\mathbf{q}$ to quantify the weight of each entropons. 
This observable is nearly $\mathbf{q}$-independent for $\tau\omega_E\lesssim 1$ but increases with $\tau\omega_E$: The larger $\tau\omega_E$ the larger is the contribution of each entropon to the entropy production. This linear behavior is due to the increase of the prefactor $T_a \sim \tau$.
For $\tau\omega_E\gtrsim1$, the value of $s(\mathbf{q}) T/(v_0^2 \gamma)$ strongly depends on $\mathbf{q}$ displaying higher values for smaller $\mathbf{q}$. In addition, $s(\mathbf{q}) T/(v_0^2 \gamma)$ decreases as $\tau\omega_E$ is increased and, consequently, shows a non-monotonic behavior with a maximum that shifts for larger $\tau\omega_E$ as $\mathbf{q}$ is decreased.

To shed light on this non-monotonicity, the height of the peak of the rescaled spectral entropy production, $\sigma(\omega^*, \mathbf{q}) T/T_a$, is shown in Fig.~\ref{fig:Fig1}~(g) versus $\tau\omega_E$ for different values of $\mathbf{q}$, revealing approximately the profile $\sim 1/(1+b \tau^2\omega_E^2)$ where $b=b(\mathbf{q})$ is a fitting parameter.
The main contribution to $\sigma(\omega^*, \mathbf{q}) T/T_a$ is due to frequencies lower than $\omega^*(\mathbf{q})$.
Considering formula~\eqref{eq:spectral_entropyprod}, the heigth of the peak of $\sigma$ is roughly determined by $\sigma(\omega^*, \mathbf{q}) T/T_a\sim K(\omega^*)$. 
By approximating $\omega \sim \omega(\mathbf{q})$, we obtain
\begin{equation}
\label{eq:Kespression_xi}
\sigma(\omega^*, \mathbf{q}) \frac{T}{T_a}= \frac{1}{1+\omega^2(\mathbf{q})\tau^2} \approx \frac{1}{1+\left(1+\frac{\tau}{\tau_I}\right)\xi^2 \mathbf{q}^2}
\end{equation}
where $\xi$ is the correlation length of the spatial velocity correlation, $\langle \mathbf{v}(\mathbf{r}) \cdot\mathbf{v}(0) \rangle$, of an active solid and reads~\cite{caprini2021spatial}
\begin{equation}
\label{eq:correlationlegnth_spatialvelocitycorrelation}
\xi^2= \frac{3}{2} \bar{x}^2 \frac{\tau^2 \tau_I}{\tau_I+\tau} \omega^2_E \,.
\end{equation}
Evaluating the denominator of Eq.~\eqref{eq:Kespression_xi}, and requiring that the $\mathbf{q}$-dependence is negligible, we determine the typical wave vectors at which $\sigma T/T_a$ starts decreasing
\begin{equation}
\mathbf{q}_*^2 = \frac{1}{\xi^2 \left( 1+\tau/\tau_I \right)}\,.
\end{equation}
The wave-vectors with $\mathbf{q} \lesssim \mathbf{q}_*\sim 1/\xi$ provide the main contribution to the rescaled entropy, while those with $\mathbf{q} \gtrsim \mathbf{q}_*\sim1/\xi$ have a much smaller weight.
Since $\xi$ increases with $\tau\omega_E$, the modes with larger $\mathbf{q}$ give a smaller contributions as $\tau\omega_E$ is increased.

\begin{figure}[!t]
\centering
\includegraphics[width=0.95\linewidth,keepaspectratio]{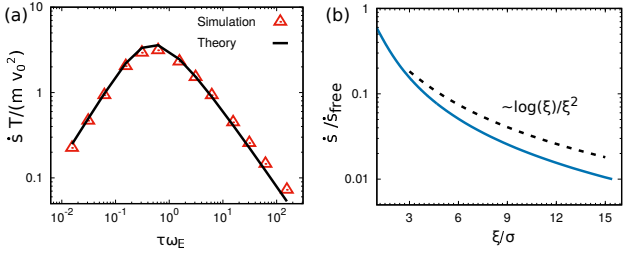}
\caption{Total entropy production. 
(a): Rescaled entropy production $\dot{s} \,T/ (mv_0^2)$ as a function of $\tau \omega_E$. 
(b): Entropy production $\dot{s}/ \dot{s}_{\text{free}}$ as a function of $\xi/d_0$.
Solid lines are obtained by theoretical predictions, while points by numerical simulations.
}\label{fig:Fig2}
\end{figure}

By integrating our prediction for $\sigma(\omega,\mathbf{q})$ (Eq.~\eqref{eq:def_spectralentropyprod}) over $\omega$ and $\mathbf{q}$, we can derive analytically the expression for the global density of entropy production of the solid, $\dot{s}$. 
\begin{equation}
\label{eq:entropyproduction_perfectsolid}
\dot{s}\approx 
\dot{s}_{\text{free}} \int_{\Omega} \frac{d{\mathbf{q}}}{\Omega}  \frac{1}{1+ \frac{\tau^2\tau_I}{\tau+\tau_I} \omega^2(\mathbf{q})}
\end{equation}
where $\dot{s}_{\text{free}}= \rho\frac{T_a}{T }(\tau_I+\tau)^{-1}$ is the density of entropy production of non-interacting active particles. $\dot{s}_{\text{free}}$ is proportional to the ratio between the active temperature ($T_a= v_0^2 \gamma \tau$) and the thermal temperature ($T$), and is a function of the dynamical properties of the system, such as persistence time $\tau$ and inertial time $\tau_I$.
The integral in Eq.~\eqref{eq:entropyproduction_perfectsolid} can be analytically expressed in terms of elliptic functions in terms of the parameters of the model (see SM).
Figure~\ref{fig:Fig2}~(a) plots $\dot{s}$ as a function of $\tau \omega_E$ showing a non-monotic behavior: in the small persistence regime ($\tau \omega_E \ll 1$), $\dot{s} \sim 0$ because the system behaves as an inertial solid in equilibrium. Increasing $\tau \omega_E$, the system departs from equilibrium and entropy production grows until reaching a maximum, roughly at $\tau \omega_E \sim 1$.
For further increase of $\tau \omega_E$, the value of $\dot{s}$ decreases until almost vanishes.

While the increase of $\dot{s}$ is expected when the system departs from equilibrium and is well-explained by the increase (up to saturation) of the non-interacting entropy production $\dot{s}_{\text{free}}\sim \tau/(\tau+\tau_I)$, the physical picture behind the decrease in the large persistence regime can be understood by expressing $\dot{s}$ as a function of $\xi$.
Here, we report the scaling behavior of $\dot{\sigma}$ for $\xi \gg 1$
\begin{equation}
\label{eq:EntropyProd_scaling}
\frac{\dot{s}}{\dot{s}_{\text{free}}} \sim 
        \dfrac{\log{\left(  \xi \right)}}{2\sqrt{3} \,\xi^2}, \text{for } \xi \gtrsim 1 \,,
\end{equation}
which is shown in Fig.~\ref{fig:Fig2}~(b).
The decrease of $\dot{s}$  as $\xi$ increases suggests that the onset of spatial velocity correlations characterizing the solid at higher densities  reduces the entropy production.
Coherent domains of strongly correlated velocities produce less entropy, 
while the incoherent behavior of the velocities of the particles 
determines a higher dissipation, as if coherent motion, somehow, minimizes the effective friction between different particles.

In conclusion, we have predicted new elementary vibrational excitations in active solids, termed entropons, which emerge from the spectral entropy production and are generated uniquely by activity. Our combined numerical and theoretical study revealed the properties of entropons and demonstrated that they dominate over phonons far from equilibrium.

The concept of ``entropons'' as additional lattice vibrations has a broad generality that goes beyond monodisperse active crystals. It will certainly apply to binary crystals composed of active and passive particles~\cite{ni2014crystallizing}.
Moreover, we expect that in disordered dense systems, such as active glasses~\cite{flenner2016nonequilibrium, nandi2018random, klongvessa2019active, paoluzzi2022motility} and dense active liquid crystals~\cite{doostmohammadi2018active}, entropons could play the dominant role of system excitations in determining entropy production. For instance, they could shed light on the activity-induced shift of the glass transition temperature. 

Many experimental realizations of active crystals are available. Examples include biological tissues~\cite{vilfan2005oscillations}, confluent cell monolayers~\cite{garcia2015physics, henkes2020dense}, dense assemblies of active colloids~\cite{klongvessa2019nonmonotonic} as well as highly packed active granular systems~\cite{dauchot2005dynamical, briand2016crystallization}, for which the solid structure has been recently achieved by connecting Hexbug particles by springs~\cite{baconnier2021selective}. Therefore the existence of entropons can in principle be verified by analyzing particle trajectories in real space.

Future studies on entropons can be manifold by including the scattering of entropons near crystalline defects, the shifted spectra of entropons in a crystal exposed to a temperature gradient. This could lead to possible applications like shock absorbers and active mass dampers as well as controlled heat radiators obtained by active crystals.


\begin{acknowledgments}
\textit{Acknowledgments --- } 
LC acknowledges support from the Alexander Von Humboldt foundation.
UMBM  and AP acknowledge support from the MIUR PRIN 2017 project
201798CZLJ. 
HL acknowledge support by the Deutsche Forschungsgemeinschaft (DFG) through the SPP 2265 under the grant number LO 418/25-1.
\end{acknowledgments}

\bibliographystyle{apsrev4-1}

\bibliography{EP}

\end{document}